\documentclass[pdflatex,sn-basic]{sn-jnl}

\usepackage{graphicx}%
\usepackage{multirow}%
\usepackage{amsmath,amssymb,amsfonts}%
\usepackage{amsthm}%
\usepackage{mathrsfs}%
\usepackage[title]{appendix}%
\usepackage{xcolor}%
\usepackage{textcomp}%
\usepackage{manyfoot}%
\usepackage{booktabs}%
\usepackage{algorithm}%
\usepackage{algorithmicx}%
\usepackage{algpseudocode}%
\usepackage{listings}%
\usepackage{bookmark}
\usepackage{anyfontsize}
\usepackage{natbib}

\renewenvironment{table}[1][]%
{\tableorg[#1]%
\tablebodyfont%
\renewcommand\footnotetext[2][]{{\removelastskip\vskip3pt%
\let\tablebodyfont\tablefootnotefont%
\hskip0pt\if!##1!\else{\smash{$^{##1}$}}\fi##2\par}}%
}{\endtableorg}

\usepackage{float}

\begin{document}

\title[Article Title]{Mitigating Consequences of Prestige in Citations of Publications}

\author*[1]{\fnm{Michael} \sur{Balzer}}\email{michael.balzer@uni-bielefeld.de}

\author*[2]{\fnm{Adhen} \sur{Benlahlou}}\email{adhen.benlahlou@uni-bielefeld.de}

\affil*[1]{\orgdiv{Bielefeld University}, \orgname{Center for Mathematical Economics}, \orgaddress{\street{Universitätsstraße 25}, \city{Bielefeld}, \postcode{33615}, \state{NW}, \country{Germany}}}
\affil*[2]{\orgdiv{Bielefeld University}, \orgname{Chair for Economic Theory and Computational Economics}, \orgaddress{\street{Universitätsstraße 25}, \city{Bielefeld}, \postcode{33615}, \state{NW}, \country{Germany}}}

\abstract{For many public research organizations, funding creation of science and maximizing scientific output is of central interest. Typically, when evaluating scientific production for funding, citations are utilized as a proxy, although these are severely influenced by factors beyond scientific impact. This study aims to mitigate the consequences of the Matthew effect in citations, where prominent authors and prestigious journals receive more citations regardless of the scientific content of the publications. To this end, the study presents an approach to predicting citations of papers based solely on observable characteristics available at the submission stage of a double-blind peer-review process. Combining classical linear models, generalized linear models and utilizing large-scale data sets on biomedical papers based on the PubMed database, the results demonstrate that it is possible to make fairly accurate predictions of citations using only observable characteristics of papers excluding information on authors and journals, thereby mitigating the Matthew effect. Thus, the outcomes have important implications for the field of scientometrics, providing a more objective method for citation prediction by relying on pre-publication variables that are immune to manipulation by authors and journals, thereby enhancing the objectivity of the evaluation process. Our approach is thus important for government agencies responsible for funding the creation of high-quality scientific content rather than perpetuating prestige.}

\keywords{Prediction, Linear Models, Negative Binomial Regression, Matthew Effect, Funding}

\maketitle

\section{Introduction} \label{sec:intro}
Every published paper is expected to contribute positively to scientific knowledge, although the significance of these contributions can vary widely. Typically, the scientific value of a paper is gauged by the number of citations it receives, which ideally reflects the quality of the research. In scientometric studies, research quality is generally assessed through factors such as the importance of the topic, strength of the evidence, novelty, study design, and methodology \citep{tahamtan2016, tahamtan2019citation, xie2022citing}. However, there is increasing evidence that various non-scientific factors also play a role in a paper's reach and impact, including stylistic choices \citep{haustein2014,letchford2016advantage, murphy2019does, heard2023if, martinez2021specialized}, the number of authors \citep{fox2016citations}, biases related to language and gender \citep{andersen2019gender}, and whether the paper is available as a preprint \citep{fu2019releasing}. The practice of using citations to evaluate the significance of scientific work dates back to 1927 when \cite{gross1927college} first applied this method. Since then, citation analysis has been widely used to assess national science policies and the development of disciplines \citep{oppenheim1995correlation, oppenheim1997correlation, lewison1998gastroenterology, tijssen2002mapping}, as well as the performance of research departments and laboratories \citep{bayer1966some, narin1976evaluative}, books and journals \citep{garfield1972citation, nicolaisen2002j}, and individual scientists \citep{garfield1970citation}. These studies typically use the number of citations of peer-reviewed papers as a measure of a scientist's impact on the scientific community, under the assumption that high-quality work generates more citations \citep{cawkell1968citation, van2003bibliometric}.

However, measuring scientific production by means of citations in the original form is unsatisfying as there is significant skepticism within the field of scientometrics about whether citations truly reflect the impact of scientific activity \citep{woolgar1991beyond, haustein2014}. These doubts trace back to \cite{garfield1972citation}, who acknowledges that citations are influenced by many variables beyond scientific impact. Therefore, scientometricians have extensively debated the non-scientific factors that can affect citations (see, for example, \cite{tahamtan2016}). These factors are vulnerable to various forms of manipulation, challenging the validity of citations as an objective metric for evaluating scientific output. For instance, self-citations, citation rings, and strategic citation placements can artificially boost citations, distorting the actual impact and quality of the research \citep{wren2022}. Moreover, the Matthew effect, where prominent authors and prestigious journals receive more citations regardless of content, further complicates the issue \citep{azoulay2014}. Consequently, citations may reflect elements other than scientific merit, such as visibility, network influence, or deliberate gaming of the system \citep{marchi2024}. Furthermore, one goal of public research organizations is to maximize research quality and not the number of citations. This diminishes their reliability as indicators of a researcher's true contributions, raising concerns about their use in evaluating scientific production, particularly when assessing the impact of collaboration on scientific output \citep{hirsch2005, nih2024}.

In this study, we propose an approach that aims at partially addressing this concerns by mitigating author and journal prestige in the citations of scientific papers. To this end, citations of papers are predicted while deliberately excluding factors related to authors and journals, which are seen as sources of bias. Although this approach generally removes any information and impact carried by the factors not considered, the main advantage is the capture and removal of the influence of the Matthew effect, where prestige disproportionately amplifies recognition and thus citations \citep{azoulay2014}. Particularly, the focus is on observable characteristics immune to manipulation which are accessible during a double-blind peer review process, such as the tendency of papers citing more recent literature to garner more citations than those referencing older work \citep{roth2012}. Additionally, we consider the correlation between citations and factors like the number of references and reference novelty \citep{ahlgren2018}. By utilizing large-scale data sets on biomedical papers based on the PubMed database in combination with estimation in the framework of classical linear models (LM) and generalized linear models (GLM), we show that high prediction accuracy can be achieved by variables exclusively available in double-blind peer review processes. Generally, the results are in line with previous literature on citation prediction, while excluding information related to authors and journals. Thus, our findings suggest that factors beyond the journal's and author's control influence a significant proportion of the citations, which indicate that reliable prediction based on pre-publication variables can be achieved. Therefore, our approach is especially important for government agencies responsible for evaluating the scientific output they fund, as one of their primary interest lies in supporting the creation of high-quality scientific content rather than perpetuating prestige. By removing prestige-related biases, our predicted citations offer a more objective evaluation of scientific output, aligning more closely with one goal of public research funding \citep{hirsch2005, nih2024}.

The structure of this paper is as follows: We provide a comprehensive literature review of related work on citation prediction to differentiate our work from previous research in Section \ref{sec:litreg}. Afterward, the methodology in this paper as well as the empirical setting is explained in detail in Sections \ref{sec:emp} and \ref{sec:meth}. The results for the empirical setting can then be viewed in Section \ref{sec:res} while this paper finalizes with a discussion and conclusion in Section \ref{sec:con}.

\section{Literature Review} \label{sec:litreg}
In this work, our main focus is on mitigating the consequences of the Matthew effect and thus author and journal related prestige by predicting the citations of papers. Therefore, we review related work on citation prediction. Given the significant role citations play in evaluating the quality of research and researchers, it is important to understand why some papers attract more citations than others. Numerous studies have explored the factors influencing citations, with some even attempting to predict the future citation impact of research \citep{garfield1970citation}.

\cite{tahamtan2016} conducted a comprehensive review of the factors that influence citation frequency, identifying 198 relevant studies. They categorized these factors into three main groups: those related to the paper itself, those related to the journal, and those related to the author. Given that the latter two categories can be influenced by prestige, they fall outside the scope of our analysis. Focusing on paper-related factors, previous research has demonstrated that citation counts are significantly impacted by the paper's quality \citep{buela2010, patterson2009, stremersch2007}, the characteristics of the field, subfield, or study subject \citep{glanzel2003, glanzel2014, dorta2014, gonzalez2016}, the nature of the references \citep{antoniou2015, biscaro2014, didegah2013, onodera2015, yu2014}, and the length of the paper \citep{falagas2013, stremersch2015, van2014}. To our knowledge, no studies have exclusively focused on observable characteristics of the manuscript that a referee would assess in a double-blind review process. The closest related work is by \cite{lokker2008prediction}, who predicted citation counts within a two-year horizon based on information available within three weeks of publication.

In scientometric research, especially in areas such as citation analysis, patent analysis, webometrics, and altmetrics, the dependent variables are typically represented as count data \citep{ajiferuke2015modelling}. Examples of such count variables in the literature include the number of citations, the research output of a scientist measured by the total publications, and the number of references to scientific literature made by patents, among various other count-based measures examined in these studies. As noted by \citep{ajiferuke2015modelling}, the predominant approach in the literature for modeling count data has been the use of linear regression models \citep{ajiferuke2005inter, ayres2000determinants, bornmann2007multiple, habibzadeh2010shorter, landes2000citations, lokker2008prediction, peters1994determinants, vaughan2005modeling, willis2011predictors, xia2011multiple, yoshikane2013multiple, yu2014}. While negative binomial regression models hold strong statistical potential, their application in scientometrics remains limited, with only a small number of studies investigating their use in this area \citep{ajiferuke2015modelling, baccini2014crossing, chen2012predictive, didegah2013, lee2007depth, mcdonald2007understanding, thelwall2015scholarly, walters2006predicting, yoshikane2013multiple, yu2014patent}.

\section{Empirical Setting} \label{sec:emp}
The empirical setting is concerned with citations of publications in the field of biomedicine. This is mainly motivated by the fact that double-blind peer review processes are the usual practice in biomedical sciences \citep{miller2024}. Furthermore, extensive observable characteristics of a paper can be recovered at the submission stage which are generally immune to manipulation from outside sources. Additionally, by combining many different data sources on publications in biomedicine, large-scale data-sets with many variables can be constructed. 

Particularly, The main dataset has been obtained from the PubMed Knowledge Graph (PKG), an extensive and open-access resource that enriches PubMed with detailed and fine-grained information \citep{xu2020building}. Similar to \cite{li2022}, this study utilizes the PKG version from the PubMed 2021 baseline. In general, the version is freely accessible from the official website \footnote{\url{http://er.tacc.utexas.edu/datasets/ped}} and contains information on more than 37 million papers. For each paper, the PKG provides comprehensive bibliographic details (such as PMID, article type, title, keywords, MeSH terms, abstract, disambiguated authors, funding information, and author affiliations), accurately extracted and normalized bio-entities (including diseases, chemicals/drugs, proteins/genes, species, and mutations), as well as its citation relationships with other PubMed papers \citep{xu2020building}.

Furthermore, the Office for Portfolio Analysis within the U.S. National Institutes of Health publishes periodically the so-called National Institutes of Health Open Citation Collection (NIH-OCC). The NIH-OCC consists of bibliographical and citation data which is specifically designed for the biomedical domain. Particularly, the NIH-OCC provides information related to the citations of articles available in the PubMed database. The dataset is additionally further divided into the Open Citation Collection concerned with the citation links between different articles and the iCite metadata, which contains metadata of bibliographic resources, including citation information. In general, the NIH-OCC is freely available in FigShare \footnote{\url{https://figshare.com/collections/The_NIH_Open_Citation_Collection_A_public_access_broad_coverage_resource/4693772}} which is an open-access repository for research materials. The separate data on iCite can also be accessed via the iCite web service \footnote{\url{https://icite.od.nih.gov}} \citep{hutchins2019nih}.

Unfortunately, the data sets have generally diverging number of observations such that comparability can only be ensured by combining both data sets based on the involved papers by the unique identifier number from PubMed (PMID). Based on the combined data set of around 30 million papers, two separate data sets are constructed which are governed by the response variable of interest. Particularly, the \textit{Citations} variable describes simply the citation counts. The corresponding data set consists of 14,471,025 observations covering the years 1998 to 2018. The large reduction in observations from the original data set occurs due to the construction of the \textit{SJR} variable which is the continuous counterpart of citation counts for comparability reasons. The motivation for the construction stems from the idea that not all citations hold equal value, especially in the realm of biomedical science. For instance, being cited by an article published in a high-tier journal provides a paper with far greater visibility and recognition than being referenced by a lower-tier journal \citep{kravitz2011}. This distinction highlights the importance of considering not just the number of citations, but the quality of the citing sources. Bibliometric indicators increasingly account for the heterogeneous nature of citations, as studies have shown that raw citation counts often fall short in reflecting the true scientific merit of a paper \citep{garfield1979citation}. By weighting citations according to the impact factor of the citing journal, a more refined measure of a paper’s value can be achieved. High-impact journals, which are generally more selective, lend greater weight to the citations they provide, signaling stronger recognition within the scientific community. To implement this approach, the SCImago Journal Rank impact factor is utilized, which captures both the quality and reputation of journals, providing a robust method for citation count weighting. The impact factor can reliably only be recovered for the years 1998 to 2018, thereby resulting in a data set of 12,755,581 observations excluding papers for which the corresponding journal does not have a SCImago Journal Rank impact factor \citep{falagas2008, roldan2019}.

A comprehensive collection of independent variables with data type and description can be seen in Table \ref{tab:var}. Generally, all variables occur in both separate data sets and amount to 106 unique independent variables. Medical subject headings (MeSH) refers to the vocabulary thesaurus for indexing articles on PubMed. Thus, the \textit{Scores} are defined as the proportion of the MeSH terms belonging to the major research domains: A (animal), C (molecular), and H (human) \citep{lowe1994}.

\begin{table}[!htpb]
\caption{\label{tab:var} Data type and description of independent variables.}
\centering
\begin{tabular}{@{\extracolsep{5pt}}lll}
\\[-1.8ex] \hline
\hline \\[-1.8ex]
\textbf{Variable} & \textbf{Type} & \textbf{Description} \\ 
\hline \\[-1.8ex]
MeSH & Numeric & Number of MeSH terms \\ 
Scores & Numeric & Proportion of MeSH terms: A, C and H \\  
Title & Numeric & Length of the title in words \\  
References & Numeric & Number of references \\ 
Age & Numeric  & Mean age of references in years \\ 
Length & Numeric  & Length of the paper in pages \\
Triangle & Categorical & Position in the biomedical triangle (see \cite{weber2013identifying}) \\ 
Year & Categorical  & Year of publication \\
Language & Categorical  & Rank of languages (1: English, 2: English/Other, 3: Other) \\ 
Clinical & Binary &  Clinical paper (Yes/No) \\ 
Research & Binary  &  Research article (Yes/No) \\
Access & Binary  & Open access (Yes/No) \\
Publication & Binary  &  Type of publication \\
\hline \\[-1.8ex]
\end{tabular}
\end{table}

Controlling for field specific citations pattern, requires a classification system in which publications are assigned to fields. To do so, we rely on the so-called biomedical triangle. In order to get a finer division of the Triangle of Biomedicine to control for field citations rate difference, we divided the main triangle into smaller subtriangles based on key geometric features. The Triangle of Biomedicine is defined by three vertices representing major research domains. To create subdivisions, we first identify the midpoints of the triangle's sides, specifically the midpoints between A and C, A and H, and C and H. These midpoints, along with the origin of the triangle (geometrically centered at O), are used to form additional subtriangles within the original triangle. For additional in-depth details and intuition for the Triangle of Biomedicine, we refer to the original work of \cite{weber2013identifying}.

Furthermore, the data sets includes information on the languages utilized in the papers. Since papers can be purely in English, bilingual or any other language, a ranking is created based on the perceived importance in the scientific community \citep{meneghini2007,van2012}. Specifically, English is assigned the rank one, bilingual papers the rank two and every paper in any other language the rank three. The ranking approach is motivated by practical challenges in statistical modeling. If every language is considered as an independent variable, the set of variable will increase substantially. By utilizing the ranking approach, the language variable is condensed to three independent variable while maintaining important information for interpretation \citep{fahrmeier2013}.

The \textit{Access} variable is constructed based on the PMC Open Access Subset provided by the National Institutes of Health. Particularly, around two million papers are matched based on PMID and thus are assigned as open access availability \citep{pmc2014}. Moreover, papers can be assigned to multiple publication types such that each publication type is treated as a binary variable indicating absence or presence thereof.

\section{Methodology} \label{sec:meth}
The general methodology considered in this work is applicable to any fields of science beyond biomedicine where double-blind peer review processes are practiced and extensive observable characteristics can be recovered. The focus is on classical statistical models, specifically LM and GLM. In this context, key issues emerging in the model-building process have to be addressed. Generally, there is a wide range of potential statistical models available for predicting citations (see, for example, \cite{ajiferuke2005inter, ajiferuke2015modelling}).

The classical LM is a versatile and powerful tool for estimation when the response variable is continuous and approximately normally distributed. The \textit{SJR} is a continuous variable which is by definition always positive and potentially equal to zero, thereby posing practical challenges when a LM is chosen in the estimation procedure since approximate normality cannot be ensured \citep{fahrmeier2013}. Furthermore, the usual log transformation cannot be utilized since the logarithm of zero is not defined. As an alternative, the inverse hyperbolic function $arsinh(\cdot)$ can be utilized. In general, the function is well-defined around values of zero and the behavior is comparable to $\log(\cdot)$ if the values of \textit{SJR} are large \citep{chen2024logs}. Thus, the \textit{SJR} is a continuous variable which approximate normality can be ensured by the $arsinh(\cdot)$ transformation as seen in Appendix \ref{sec:app1}.
Under the assumption that the \textit{SJR} is normally distributed
\begin{equation*}
    SJR_i \sim N(\mu_i, \sigma^2)
\end{equation*}
the following model formulation can be utilized
\begin{equation*}
    E(\textit{SJR}_i) = \mu_i = \mathbf{x'_i}\boldsymbol{\beta_i}
\end{equation*}
such that $i \in \{1, \cdots, n\}$, where $n$ is the number of observations, $\mathbf{x'_i}$ are the independent variables of interest and $\boldsymbol{\beta_i}$ corresponding coefficients. Generally, we must be cautious in the interpretation of the coefficients due to the nature of the inverse hyperbolic transformation. Specifically, for large values of the independent variables, the effect on the response variable is approximately similar to that in a log-log model, where a 1\% increase in the variable ceteris paribus leads to an approximate $\beta_i\%$ increase in $\mu_i$ on average. In contrast, for small values of the independent variables, the interpretation resembles that of a linear model, where a one-unit increase in the variable leads to an approximate increase of $\beta_i$ units in $\mu_i$, holding other variables constant, on average. Thus, depending on the magnitude of the continuous variables, the effect on $\mu_i$ can vary between these two interpretations \citep{chen2024logs}.

While it is possible to treat citation counts as a continuous variable and ensure approximate normality trough appropriate transformations, applying classical LM overlooks the count-based nature of the data. Citation counts are additionally often zero-inflated and overdispersed as seen in Appendix \ref{sec:app1}, which classical LM cannot adequately \citep{ajiferuke2015modelling}. To address these challenges, we explore appropriate statistical models, accounting for the specific distributional characteristics of citations count data. In general, the unifying framework of GLM allows for a flexible estimation even when the response variable is not normally distributed \citep{fahrmeier2013}. To this end, the mean of the response variable is connected to the linear predictor $\eta_i = h(\mathbf{x'_i}\boldsymbol{\alpha_i})$, where $\mathbf{x'_i}$ are the independent variables of interest as before and $\boldsymbol{\alpha_i}$ the corresponding coefficients in the GLM, via the response function $h(\cdot)$. Furthermore, the probability density function of the response variable can be rewritten in the form of the univariate exponential family \citep{nelder1972, mccullagh1989}. However, the class of statistical models for dealing with count data is large, making the choice for an appropriate model an important task. It has been noted by \cite{zeileis2008} that in the presence of overdispersion and zero-inflation, negative binomial regression models are promising candidates for estimation \citep{mccullagh1989}. Under the assumption that the citation counts on the original scale are negative binomial distributed 
\begin{equation*}
    \textit{Citations}_i \sim NB(\nu_i, \psi),
\end{equation*}
the expectation parameter $\nu_i$ is modeled via the linear predictor with exponential response function as
\begin{equation*}
   E(\textit{Citations}_i) = \nu_i = h(\mathbf{x'_i}\boldsymbol{\alpha_i}) = \exp(\mathbf{x'_i}\boldsymbol{\alpha_i}).
\end{equation*}
Regarding the interpretation of the effects of independent variables on the $\nu_i$, we have to consider the multiplicative effect induced by the response function. Thus, an increase in any continuous variable by one unit ceteris paribus leads to a $100 \cdot (\exp(\alpha_i) - 1)$ percent increase in $\nu_i$ on average \citep{fahrmeier2013}. 

Another key issue emerges as the dataset contains a vast array of explanatory variables, making the selection of potentially relevant factors a complex task \citep{fahrmeier2013}. This requires a combination of scientometrician's expertise and careful consideration of the underlying data structure to identify the most meaningful variables. The variables considered in the models include, but are not limited to, the year of publication, number of references, publication type, and language. The primary goal of our approach is to develop a tailored prediction model for each paper, focusing on variables observable during a double-blind peer review process. By focusing on these characteristics, we aim to mitigate the influence of the Matthew effect, where more prestigious authors or journals tend to receive disproportionate recognition \citep{azoulay2014}. Additionally, parsimonious parametric models are derived from the complete model formulation. Specifically, we refer to the model incorporating all variables as the non-parsimonious, complete linear model (LMC). In contrast, the models including only iCite metadata and only numeric variables are referred to as the iCite metadata model (LMI) and the numeric variables model (LMR), respectively. Analogous model building processes are conducted for the GLM, where the model incorporating all variables is referred to as the non-parsimonious, complete GLM (GLMC), iCite metadata model (GLMI) and numeric variables model (GLMR) \citep{hutchins2019nih}. In Appendix \ref{sec:app3}, we additionally introduce model-based gradient boosting as a machine learning technique as an alternative estimation procedure in the framework of LM and GLM with inherent data-driven variable selection beyond the scientometrician's expertise for robustness checks \citep{hepp2016}.

\section{Results} \label{sec:res}
In the following sections, we will apply the methods introduced in Section \ref{sec:meth} to our empirical setting. The results are detailed in two subsections, where the estimated coefficients are presented for the LM followed by the GLM. Results for the alternative machine learning procedure with data-driven variable selection are presented in Appendix \ref{sec:app3}. To evaluate the predictions, we estimate the statistical models based on a train data set, which is constructed by randomly selecting 80\% of papers published in a year. Consequently, the leftover 20\% are aggregated in a test data set. Furthermore, we provide additional robustness checks by providing additional results for a 50\% split in the Appendix \ref{sec:app2}. For train and test data set, the performance is evaluated based on typical criteria. In particular, the negative log-likelihood (NNL) based on the probability density functions is computed for each model \citep{fahrmeier2013}. Furthermore, the Akaike information criterion (AIC) defined as 
\begin{equation*}
    \text{AIC} = 2k - 2 \log L
\end{equation*}
where $L$ is the maximized log-likelihood and $k$ the number of coefficients is compared \citep{akaike1974}. Similarly, the Bayesian information criterion (BIC) is found via 
\begin{equation*}
\text{BIC} = \log(n_{\textit{train}}) \cdot k - 2 \log L
\end{equation*}
where $n_{\textit{train}}$ is the number of observations in the train data set \citep{vrieze2012}. For the classical LM, the standard $R^2$ is additionally reported for the train data set. Common performance criteria for the evaluation of the prediction on test data set are the mean squared error of prediction (MSEP) defined as
\begin{equation*}
    \text{MSEP} = \frac{1}{n_{\textit{test}}} \sum_{i=1}^{n_{\textit{test}}} (y_i - \hat{y}_i)^2
\end{equation*}
and the mean absolute error (MAE) defined as
\begin{equation*}
    \text{MAE} = \frac{1}{n_{\textit{test}}} \sum_{i=1}^{n_{\textit{test}}} \left| y_i - \hat{y}_i \right|
\end{equation*}
where $n_{\textit{test}}$ is the number of observations in the test data set and $y_i$ the respective response variable of interest. For all proposed performance criteria, lower values are always preferred \citep{fahrmeier2013}. The estimation of LM and GLM is performed in the programming language R \citep{R}. The R code and data for reproducibility of all results is freely available at \url{https://doi.org/10.5281/zenodo.14050161}. 

\subsection{Predicting Weighted Citations with Linear Models} 
In this section, we will focus on the evaluation of the prediction of the weighted citations with the classical LM based on the constructed models from Section \ref{sec:meth}. For better readability and interpretability, we drop the variables controlling for the year of the publication as well as the publication type in the presentation of the results. The estimation results for the LM can be seen in Table \ref{tab:ols}.

\begin{table}[!htbp]
\centering
\caption{Estimation Results of weighted citation on train data set via linear models} 
\label{tab:ols} 
\begin{tabular}{lccc} 
\\[-1.8ex] \hline
\hline \\[-1.8ex]
\textbf{Variable} & \textbf{LMC} & \textbf{LMI} & \textbf{LMR} \\ 
\hline \\[-1.8ex]
Clinical & 0.066$^{***}$ & 0.363$^{***}$ &  \\ 
Research & 0.296$^{***}$ & 0.178$^{***}$ &  \\ 
H Score & $-$0.453$^{***}$ & $-$0.521$^{***}$ & $-$0.781$^{***}$ \\ 
A Score & $-$0.305$^{***}$ & $-$0.332$^{***}$ & $-$0.455$^{***}$ \\ 
C Score & 0.253$^{***}$ & 0.195$^{***}$ & 0.105$^{***}$ \\ 
Title & $-$0.040$^{***}$ & $-$0.009$^{***}$ & $-$0.066$^{***}$ \\ 
Triangle ACH  & 0.226$^{***}$ & 0.202$^{***}$ &  \\ 
Triangle C  & 0.018$^{***}$ & $-$0.005 &  \\ 
Triangle H  & 0.038$^{***}$ & 0.014$^{***}$ &  \\ 
Triangle Outside  & 0.126$^{***}$ & 0.098$^{***}$ &  \\ 
References & 0.713$^{***}$ & 0.753$^{***}$ & 0.722$^{***}$ \\ 
Age & $-$0.511$^{***}$ & $-$0.548$^{***}$ & $-$0.480$^{***}$ \\ 
MeSH & 0.134$^{***}$ & 0.131$^{***}$ & 0.227$^{***}$ \\ 
Access & 0.082$^{***}$ & 0.127$^{***}$ &  \\ 
Language 2 & $-$0.830$^{***}$ & $-$0.860$^{***}$ &  \\ 
Language 3 & $-$1.370$^{***}$ & $-$1.395$^{***}$ &  \\ 
Length & 0.062$^{***}$ & 0.117$^{***}$ & 0.116$^{***}$ \\ 
Constant & 1.935$^{***}$ & 2.038$^{***}$ & 1.760$^{***}$ \\ 
\hline 
Observations & 10204457 & 10204457 & 10204457 \\ 
Residual Std. Error & 1.277  & 1.291  & 1.408  \\ 
F Statistic & 72,522.600$^{***}$  & 195,640.100$^{***}$ & 556,534.800$^{***}$ \\ 
\hline
\hline
\textit{Note:} & \multicolumn{3}{r}{$^{*}$p$<$0.1; $^{**}$p$<$0.05; $^{***}$p$<$0.01} 
\end{tabular} 
\end{table}

Our analysis reveals that the coefficients in our models are highly significant, as evidenced by their very low p-values. Additionally, the F-statistics for all models is also highly significant. While the magnitudes of the coefficients vary across different models, their algebraic signs remain consistent. Furthermore, the general findings are consistent with established scientometric literature on biomedicine \citep{li2022}. Specifically, the number of references, MeSH terms, and the length of the paper all have a significant positive effect on weighted citations. For instance, a $1\%$ increase in the number of references ceteris paribus results in a $0.722\%$ increase in weighted citations for papers with a high number of references on average. Conversely, for papers with fewer references, adding one more reference increases weighted citations by an average of $0.722$ units, holding other variables constant. The mean age of references negatively impacts weighted citations, indicating that papers citing older sources are cited less. The length of the title has a minor negative effect. Notably, while the proportion of human and animal MeSH terms negatively affects weighted citations, the proportion of molecular MeSH terms has a positive impact. In the larger, non-parsimonious models involving the Triangle of Biomedicine, similar patterns are observed. The inclusion of additional binary variables confirms that papers focused on clinical research generally receive higher weighted citations, with an average increase of $0.363$ units or $0.363\%$ for clinical research papers compared to those on other topics. Additionally, our analysis shows that papers that are at least bilingual or not in English tend to have fewer weighted citations on average.

Finally, we evaluate the performance criteria of the LM on the train as well as test data. The results can be seen in Table \ref{tab:perf}. In all models, the train data set consists of $10204457$ observations making the models comparable. The parsimonious parametric model LMR has a R$^{2}$ of $0.303$ which shows that these eight variable are enough to explain around $30\%$ of the variation in the response variable. Additionally including the binary and categorical variables leads to an improvement in the R$^{2}$ to $0.415$. Thus, additionally around $10\%$ of the variation in the response variable can be explained by the language of the publication and if the paper is concerned with research or a clinical topics. Adding binary variables for the publication type yields only a negligible improvement in the R$^{2}$ by $0.012$. The adjusted R$^{2}$ remains identical to the standard R$^{2}$ indicating that the R$^{2}$ is not inflated by the inclusion of additional variables. Taking into account that we only investigate models with variable accessible during double-blind review processes, the results indicate that variables accessible at the submission stage of double-blind peer review process carry a large proportion of explanatory power. 

\begin{table}[!htbp]
\centering
\caption{Performance evaluation for the train and test data sets of weighted citations with linear models} 
\label{tab:perf} 
\begin{tabular}{lcccccc} 
\\[-1.8ex] \hline
\hline \\[-1.8ex]
& \multicolumn{3}{c}{\textbf{Train}} & \multicolumn{3}{c}{\textbf{Test}} \\ 
\cmidrule(r){2-4} \cmidrule(r){5-7} 
\textbf{Metric} & \textbf{LMC} & \textbf{LMI} & \textbf{LMR} & \textbf{LMC} & \textbf{LMI} & \textbf{LMR} \\ 
\hline \\[-1.8ex]
NLL & 16973532 & 17082415 & 17970447 & 4243719 & 4271119 & 4493490 \\ 
R$^{2}$ & 0.427 & 0.415 &  0.303  & - & - & - \\ 
Adj. R$^{2}$ & 0.427 & 0.415 &  0.303  & - & - & - \\ 
AIC  & 33947279 & 34164907 & 35940914 & - & - & - \\ 
BIC & 33948792 & 34165459 & 35941055 & - & - & -\\ 
MSEP &  1.630 & 1.666 & 1.982 &  1.631 & 1.666 &  1.984 \\ 
MAE & 1.006 & 1.017 &  1.122 & 1.005 & 1.017 &  1.122  \\ 
\hline \\[-1.8ex]
\end{tabular}
\end{table}

Moreover, the performance criteria based on the likelihood indicate that the model with the inclusion of all variables is the preferred option, although the difference in absolute numbers between the LMC and LMI is generally negligible. Similarly, the MSEP and MAE indicate that the best model in terms of prediction of the train data is again the LMC, followed by LMI and LMR. The results for the test data are almost identical to the train data results. We observe that the NLL for the LMC model is again best, followed by LMI and LMR. Furthermore, the MSEP and MAE are identical up to small rounding digits. This indicates that the model perform not only considerably well for the train data but also for non-observable test data. Thus, our models are generally robust and consist in terms of prediction.

\subsection{Predicting Citation Counts with Negative Binomial Regression} 
In this section, we will focus on the evaluation of the prediction of citation counts directly with GLM based on the parsimonious model constructions. For better readability and interpretability, we drop the variables controlling for the year of the publication as well as the publication type in the presentation of the results. The estimated coefficients for the GLM can be seen in Table \ref{tab:glm}.

\begin{table}[!htbp]
\centering
\caption{Estimation Results of citation counts on train data set via negative binomial regression} 
\label{tab:glm} 
\begin{tabular}{lccc} 
\\[-1.8ex] \hline
\hline \\[-1.8ex]
\textbf{Variable} & \textbf{GLMC} & \textbf{GLMI} & \textbf{GLMR} \\ 
\hline \\[-1.8ex]
Clinical & 0.096$^{***}$ & 0.466$^{***}$ &  \\ 
Research & 0.451$^{***}$ & 0.153$^{***}$ &  \\ 
H Score & $-$0.075$^{***}$ & $-$0.137$^{***}$ & $-$0.322$^{***}$ \\ 
A Score & $-$0.217$^{***}$ & $-$0.222$^{***}$ & $-$0.333$^{***}$ \\ 
C Score & 0.011$^{***}$ & $-$0.072$^{***}$ & $-$0.046$^{***}$ \\ 
Title & $-$0.069$^{***}$ & $-$0.040$^{***}$ & $-$0.029$^{***}$ \\ 
Triangle ACH  & 0.123$^{***}$ & 0.085$^{***}$ &  \\ 
Triangle C  & $-$0.078$^{***}$ & $-$0.114$^{***}$ &  \\ 
Triangle H  & $-$0.017$^{***}$ & $-$0.086$^{***}$ &  \\ 
Triangle Outside  & 0.029$^{***}$ & $-$0.031$^{***}$ &  \\ 
References & 0.576$^{***}$ & 0.625$^{***}$ & 0.630$^{***}$ \\ 
Age & $-$0.262$^{***}$ & $-$0.288$^{***}$ & $-$0.197$^{***}$ \\ 
MeSH & 0.123$^{***}$ & 0.113$^{***}$ & 0.150$^{***}$ \\ 
Access & 0.294$^{***}$ & 0.391$^{***}$ &  \\ 
Language 2 & $-$0.634$^{***}$ & $-$0.645$^{***}$ &  \\ 
Language 3 & $-$1.966$^{***}$ & $-$1.731$^{***}$ &  \\ 
Length & 0.148$^{***}$ & 0.234$^{***}$ & 0.215$^{***}$ \\ 
Constant & 1.263$^{***}$ & 1.441$^{***}$ & 0.966$^{***}$ \\ 
\hline 
Observations & 11576811 & 11576811 & 11576811 \\ 
$\psi$ & 0.803$^{***}$  & 0.768$^{***}$ & 0.667$^{***}$  \\  
\hline
\hline
\textit{Note:} & \multicolumn{3}{r}{$^{*}$p$<$0.1; $^{**}$p$<$0.05; $^{***}$p$<$0.01} 
\end{tabular} 
\end{table}

It can be seen that the coefficients as well as the nuisance parameter $\psi$ in all models are highly significant. Once again only the magnitudes of the coefficients vary across models while their algebraic sign remains consistent. Furthermore, the coefficients are also in line with the results obtained from the classical LM in terms of algebraic signs and magnitude. Thus, our results for the original citation counts is in line with established scientometric literature on biomedicine \citep{li2022}. Specifically, the number of references, MeSH terms, and the length of the paper and title all have a significant positive effect on the citation counts. For instance, a one unit increase in the number of references results in a $100 \cdot (\exp(0.630) - 1) = 87.76106$ percent increase in $\nu_i$ on average while holding the other variables constant. The interpretation remains identical for the other continuous variables. Notably, while the direction of the effects for the proportion of the MeSH terms remains, the magnitude is severely lower than in the LM, suggesting that the effect on the $\nu_i$ is much weaker. Similarly, the Triangle of Biomedicine in the larger, non-parsimonious models exhibits coefficients close to zero indicating a weak effect on the $\nu_i$. In contrast, the additional binary variables reveal similar outcomes as in the LM. Specifically, $\nu_i$ increases ceteris paribus by $100 \cdot (\exp(0.559) - 1 = 59.3607$ percent for clinical research papers on average. Additionally, the bilingual or papers that are not in English tend to have a negative impact on the $\nu_i$ on average as before.

Additionally, we evaluate the performance criteria of the negative binomial regression models on the train as well as test data. The results can be seen in Table \ref{tab:perf2}. 

\begin{table}[!htbp]
\centering
\caption{Performance evaluation for the train and test data sets of citation counts with negative binomial regression} 
\label{tab:perf2} 
\begin{tabular}{lcccccc} 
\\[-1.8ex] \hline
\hline \\[-1.8ex]
& \multicolumn{3}{c}{\textbf{Train}} & \multicolumn{3}{c}{\textbf{Test}} \\ 
\cmidrule(r){2-4} \cmidrule(r){5-7} 
\textbf{Metric} & \textbf{GLMC} & \textbf{GLMI} & \textbf{GLMR} & \textbf{GLMC} & \textbf{GLMI} & \textbf{GLMR} \\ 
\hline \\[-1.8ex]
NLL & 45488614 & 45763261 & 46653569 &  11370556 & 11438778 & 11664139\\ 
AIC  & 90977441 & 91526600 & 93307158& - & - & - \\ 
BIC & 90978968 & 91527156 & 93307301 & - & - & -\\ 
MSEP & 11013.972 &  10582.094 & 10765.797 & 15832.063 & 15591.904 & 15776.239 \\ 
MAE &  25.706 & 26.033 &  27.563 &  25.676 & 26.0153 & 27.540  \\ 
\hline \\[-1.8ex]
\end{tabular}
\end{table}

The performance criteria based on the likelihood indicate that the model with the inclusion of all variables is the preferred option, although the difference in absolute numbers between the GLMC and GLMI is generally very low. Similarly, the MAE indicates that the best model in terms of prediction of the train data is again the GLMC, followed by GLMI and GLMR. However, the MSEP reveals that the second model has the better performance on train data, indicating that the large, non-parsimonious and complete parsimonious model have problems predicting the outliers whereas the GLMI performs better. The results for the test data are almost identical to the train data results besides the magnitude in absolute values of the MSEP. We observe that the NLL for the GLMC model is again best, followed by GLMI and GLMR. This indicates that the model perform not only considerably well for the train data but also for non-observable test data, although we have to point out that the outliers in the test data heavily influence the MSEP.

Our results are in line with previous studies (see, for example, \cite{li2022}), but the novelty lies in our ability to achieve prediction quality comparable to past research using only variables that are observable pre-publication by referees in a double-blind peer-review process. This approach highlights that effective citation prediction can be conducted using pre-publication variables, making it a practical and accessible method for evaluating the potential impact of scientific papers.

\section{Conclusion} \label{sec:con}
The key findings and main contributions of this paper are: (a) The Matthew effect, where well-known authors or institutions receive disproportionate attention, regardless of the paper's merit imposes consequences in the evaluation of the contribution of scientific papers \citep{azoulay2014}. We propose to mitigate or even eliminate the Matthew effect by deliberately excluding variables related to authors and journals, thereby removing sources of bias stemming from these variables. (b) To this end, we utilize LM and GLM for the predictions of weighted citations and citation counts in the context of biomedical papers. The results show a fairly strong explanatory power on train and test data paired with highly significant coefficients. Furthermore, our results are in line with previous studies (see, for example, \cite{li2022}) with the novelty of achieving prediction quality comparable to past research with only variables accessible in double-blind peer-review processes. Thus, the potential impact of scientific papers in the context of biomedical research can be evaluated based on pre-publication variables. Importantly, the study demonstrates the feasibility of predicting the citation rate of biomedical articles using only features that are available at the time of submission. Our findings indicate that factors beyond the authors' control, such as the number of MeSH terms or publication type, explain a significant proportion of citation counts. These results suggest that useful predictions can be made without relying on post-publication data, which could lead to fairer assessments in the peer review process \citep{hirsch2005, nih2024}. (c) In the Appendices \ref{sec:app2} and \ref{sec:app3}, we provide additional robust checks by changing the split of train and test data as well as utilizing the machine learning technique called gradient boosting which has an inherent data-driven mechanism for identification of important variables and flexible model building \citep{hepp2016}. The results show that models based on gradient boosting can achieve a similar performance in terms of prediction on train and test data compared to non-parsimonious models estimated via the maximum likelihood principle. Furthermore, performance criteria and estimated coefficients are inline with the results on the 80\% split, indicating overall a robust estimation procedure.

Unfortunately, several limitations can be assessed. First, the results are presented in the context of biomedicine and therefore, only hold in that particular research area. Although we point out that the methodological approach is strictly speaking fairly general, different results might be obtained in the context of other research fields. For instance, the explanatory power of the independent variables on the variation of the response variable might vary substantially in different field of science. Furthermore, the effect of these variables indicated by the coefficients might be additionally different. Second, we utilized extensive data collection for obtaining a large set of variables. Obviously, there still might be other variables related to papers that are not considered in this study. Third, in the complete analysis only linear variables were utilized. It is realistic to assume that there might be a relationship between the response variable and the independent variables beyond the linear form. Fourth, through the exclusion of the prestige-related factors, impact and influence of variables beyond prestige are additionally excluded, indicating that targeting specifically the Matthew effect is generally a complicated endeavor \citep{azoulay2014}.

Thus, there is an outlook for certain extensions and improvements of this work. For instance, an extensive search for additional variables from different data sources can be conducted. Therefore, for future work, we plan to incorporate additional types of information, such as stylistic analysis or other text-mined features available during the submission phase, which may further improve prediction performance \citep{tahamtan2016}. Additionally, we intend to use these predictions to explore other quantitative scientometric issues. For example, citation predictions could be employed to study the effect of interdisciplinarity on citation counts, excluding network effects such as author reputation or institutional affiliation. Nevertheless, we encourage scientometricians working in the field of citations prediction to consider the consequences of the Matthew effect, if the goal is to quantify the actual scientific content of publications across all fields of science.

\begin{appendices}
\section{Additional Information on the Empirical Setting} \label{sec:app1}

\begin{figure}[H] 
    \centering
    \includegraphics[width = \textwidth]{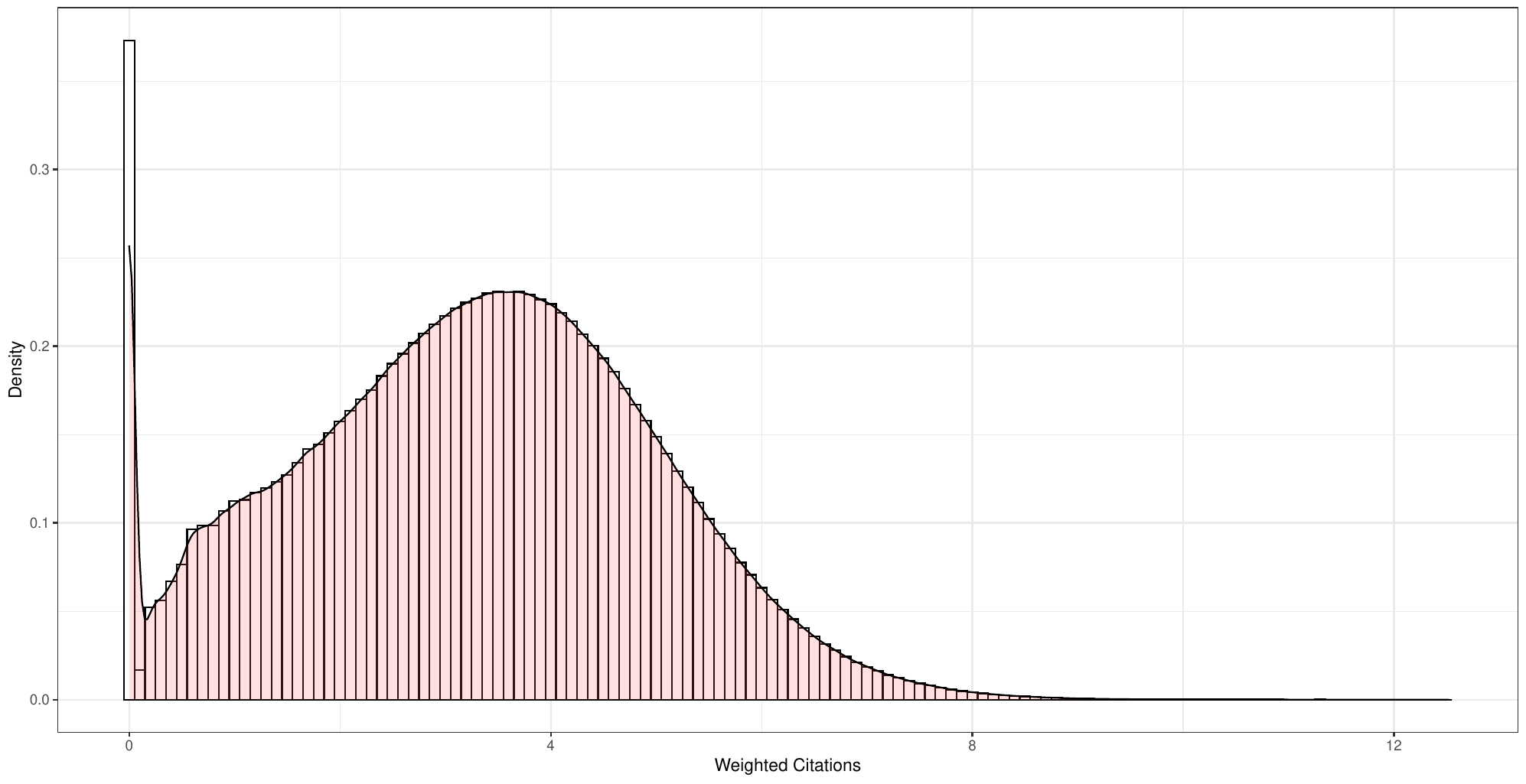}
    \caption{Histogram and kernel density of weighted citations (\textit{SJR})}
    \label{fig:emp1}
\end{figure}

\begin{figure}[H] 
    \centering
    \includegraphics[width = \textwidth]{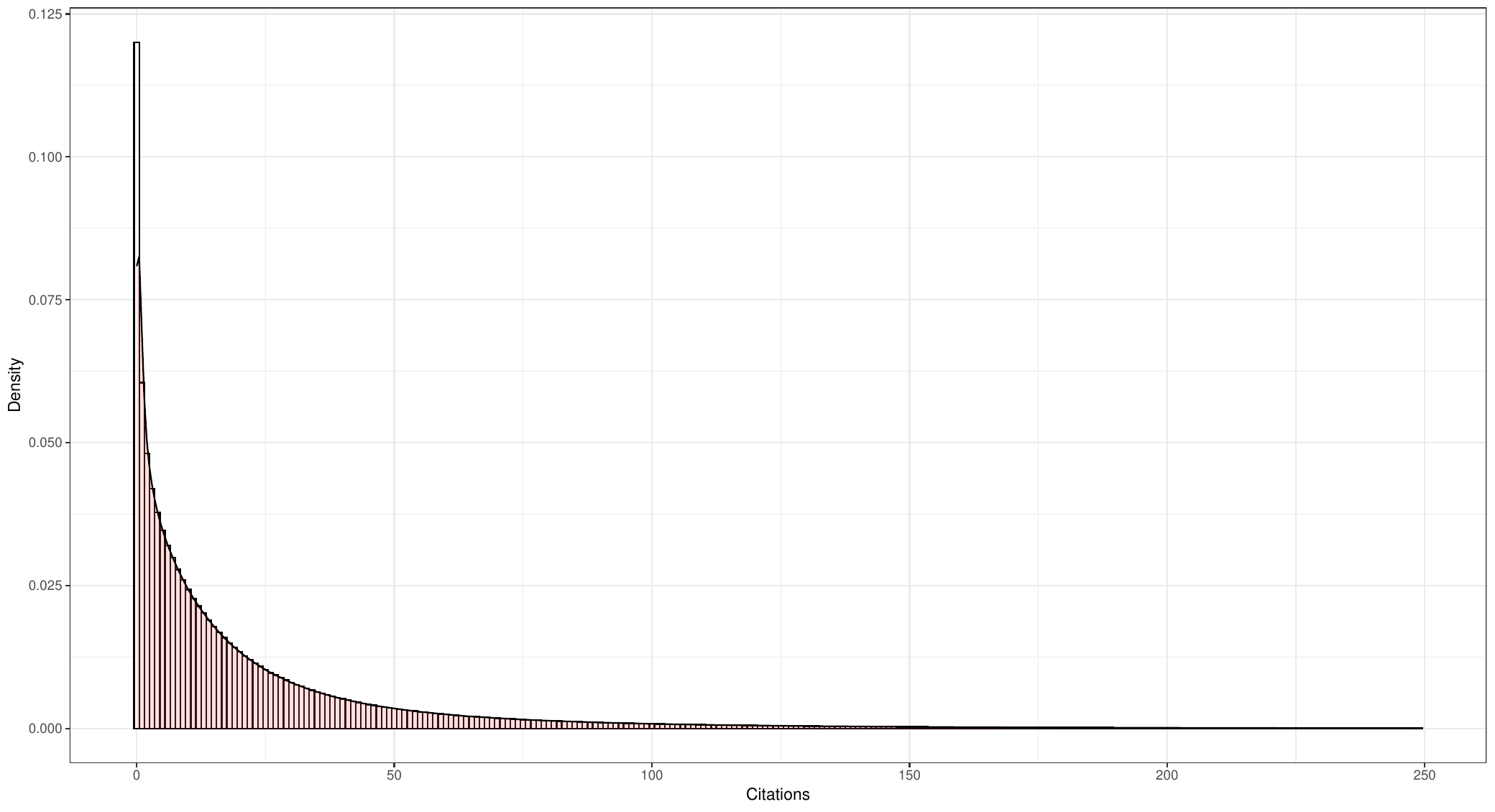}
    \caption{Histogram and kernel density of citation counts ($Citations_i \leq 250$)}
    \label{fig:emp2}
\end{figure}

\section{Results for Linear Models and Negative Binomial Regression on the 50-50 split} \label{sec:app2}

\begin{table}[H]
\centering
\caption{Estimation Results of weighted citation on train data set with $50-50$ split via linear models} 
\label{tab:olsa} 
\begin{tabular}{lccc} 
\\[-1.8ex] \hline
\hline \\[-1.8ex]
\textbf{Variable} & \textbf{LMC} & \textbf{LMI} & \textbf{LMR} \\ 
\hline \\[-1.8ex]
Clinical & 0.066$^{***}$ & 0.366$^{***}$ &  \\ 
Research & 0.288$^{***}$ & 0.177$^{***}$ &  \\ 
H Score & $-$0.454$^{***}$ & $-$0.522$^{***}$ & $-$0.779$^{***}$ \\ 
A Score & $-$0.303$^{***}$ & $-$0.330$^{***}$ & $-$0.454$^{***}$ \\ 
C Score & 0.256$^{***}$ & 0.198$^{***}$ & 0.106$^{***}$ \\ 
Title & $-$0.040$^{***}$ & $-$0.009$^{***}$ & $-$0.066$^{***}$ \\ 
Triangle ACH  & 0.229$^{***}$ & 0.206$^{***}$ &  \\ 
Triangle C & 0.021$^{***}$ & $-$0.001 &  \\ 
Triangle H  & 0.039$^{***}$ & 0.014$^{***}$ &  \\ 
Triangle Outside & 0.128$^{***}$ & 0.100$^{***}$ &  \\ 
References & 0.714$^{***}$ & 0.754$^{***}$ & 0.722$^{***}$ \\ 
Age & $-$0.511$^{***}$ & $-$0.548$^{***}$ & $-$0.480$^{***}$ \\ 
MeSH & 0.134$^{***}$ & 0.131$^{***}$ & 0.227$^{***}$ \\ 
Access & 0.080$^{***}$ & 0.125$^{***}$ &  \\ 
Language 2 & $-$0.838$^{***}$ & $-$0.870$^{***}$ &  \\ 
Language 3 & $-$1.366$^{***}$ & $-$1.393$^{***}$ &  \\ 
Length & 0.062$^{***}$ & 0.117$^{***}$ & 0.116$^{***}$ \\
Constant & 1.935$^{***}$ & 2.030$^{***}$ & 1.759$^{***}$ \\ 
\hline 
Observations & 6377785 & 6377785 & 6377785 \\ 
Residual Std. Error & 1.277  & 1.290  & 1.408  \\ 
F Statistic & 45,353.000$^{***}$  & 122,358.400$^{***}$ & 348,205.200$^{***}$  \\  
\hline
\hline
\textit{Note:} & \multicolumn{3}{r}{$^{*}$p$<$0.1; $^{**}$p$<$0.05; $^{***}$p$<$0.01} 
\end{tabular} 
\end{table}

\begin{table}[H]
\centering
\caption{Performance evaluation for the train and test data sets of weighted citations with $50-50$ split with linear models} 
\label{tab:perfa} 
\begin{tabular}{lcccccc} 
\\[-1.8ex] \hline
\hline \\[-1.8ex]
& \multicolumn{3}{c}{\textbf{Train}} & \multicolumn{3}{c}{\textbf{Test}} \\ 
\cmidrule(r){2-4} \cmidrule(r){5-7} 
\textbf{Metric} & \textbf{LMC} & \textbf{LMI} & \textbf{LMR} & \textbf{LMC} & \textbf{LMI} & \textbf{LMR} \\ 
\hline \\[-1.8ex]
NLL & 10607529 & 10675477 & 11230368 & 10609769 & 10678072 & 11233573 \\ 
R$^{2}$ &  0.427 &  0.415 & 0.304 & - & - & - \\ 
Adj. R$^{2}$ & 0.427 &  0.415 & 0.304  & - & - & - \\ 
AIC  &  21215272 & 21351033 & 22460756 & - & - & - \\ 
BIC &  21216734 & 21351566 & 22460893 & - & - & -\\ 
MSEP & 1.630 &   1.665 &  1.981 &  1.631 & 1.666 & 1.983 \\ 
MAE & 1.005 &  1.017 &  1.121 & 1.006 & 1.018 &  1.122  \\ 
\hline \\[-1.8ex]
\end{tabular}
\end{table}

\begin{table}[H]
\centering
\caption{Estimation Results of citation counts on train data set with $50-50$ split via negative binomial regression} 
\label{tab:glma} 
\begin{tabular}{lccc} 
\\[-1.8ex] \hline
\hline \\[-1.8ex]
\textbf{Variable} & \textbf{GLMC} & \textbf{GLMI} & \textbf{GLMR} \\ 
\hline \\[-1.8ex]
Clinical & 0.085$^{***}$ & 0.457$^{***}$ &  \\ 
Research & 0.467$^{***}$ & 0.155$^{***}$ &  \\ 
H Score & $-$0.077$^{***}$ & $-$0.141$^{***}$ & $-$0.317$^{***}$ \\ 
A Score & $-$0.213$^{***}$ & $-$0.219$^{***}$ & $-$0.332$^{***}$ \\ 
C Score & 0.011$^{***}$ & $-$0.072$^{***}$ & $-$0.045$^{***}$ \\ 
Title & $-$0.070$^{***}$ & $-$0.041$^{***}$ & $-$0.030$^{***}$ \\ 
Triangle ACH & 0.123$^{***}$ & 0.081$^{***}$ &  \\ 
Triangle C  & $-$0.074$^{***}$ & $-$0.107$^{***}$ &  \\ 
Triangle H  & $-$0.021$^{***}$ & $-$0.089$^{***}$ &  \\ 
Triangle Outside  & 0.027$^{***}$ & $-$0.033$^{***}$ &  \\ 
References & 0.577$^{***}$ & 0.626$^{***}$ & 0.630$^{***}$ \\ 
Age & $-$0.265$^{***}$ & $-$0.292$^{***}$ & $-$0.200$^{***}$ \\ 
MeSH & 0.123$^{***}$ & 0.113$^{***}$ & 0.149$^{***}$ \\ 
Access & 0.299$^{***}$ & 0.396$^{***}$ &  \\ 
Language 2 & $-$0.675$^{***}$ & $-$0.685$^{***}$ &  \\ 
Language 3 & $-$1.964$^{***}$ & $-$1.731$^{***}$ &  \\ 
Length & 0.150$^{***}$ & 0.235$^{***}$ & 0.216$^{***}$ \\ 
Constant & 1.261$^{***}$ & 1.449$^{***}$ & 0.969$^{***}$ \\ 
\hline 
Observations & 7,235,508 & 7,235,508 & 7,235,508 \\ 
$\psi$ & 0.803$^{***}$   & 0.768$^{***}$  ) & 0.667$^{***}$   \\ 
\hline
\hline
\textit{Note:} & \multicolumn{3}{r}{$^{*}$p$<$0.1; $^{**}$p$<$0.05; $^{***}$p$<$0.01} 
\end{tabular} 
\end{table}

\begin{table}[H]
\centering
\caption{Performance evaluation for the train and test data sets with 50-50 split of citation counts with negative binomial regression} 
\label{tab:perf2a} 
\begin{tabular}{lcccccc} 
\\[-1.8ex] \hline
\hline \\[-1.8ex]
& \multicolumn{3}{c}{\textbf{Train}} & \multicolumn{3}{c}{\textbf{Test}} \\ 
\cmidrule(r){2-4} \cmidrule(r){5-7} 
\textbf{Metric} & \textbf{GLMC} & \textbf{GLMI} & \textbf{GLMR} & \textbf{GLMC} & \textbf{GLMI} & \textbf{LMR} \\ 
\hline \\[-1.8ex]
NLL &  28425328 & 28597993 &  29155216 & 28434325 & 28604167 & 29162542 \\ 
AIC  &  56850871 & 57196064 & 58310453 & - & - & - \\ 
BIC & 56852347 & 57196602 & 58310591 & - & - & -\\ 
MSEP & 10797.267 & 10293.633 &  10474.001 & 13300.893 & 12874.068 &  13060.269 \\ 
MAE & 25.667 & 25.989 &  27.508 & 25.724 & 26.045 & 27.564  \\ 
\hline \\[-1.8ex]
\end{tabular}
\end{table}

\section{Gradient Boosting for Mitigating the Matthew Effect} \label{sec:app3}
In Section \ref{sec:intro}, we introduced the notion of deliberately excluding factors when working with citations to remove the influence of these factors. Even with the exclusion of the factors related to authors and journals, the potential number of relevant factors is still large leading to a rich set of nested statistical models. In the context of scientometrics and citation prediction, researchers are often concerned with identifying the most suitable model for predicting the impact of scientific publications \cite{fahrmeier2013}. An essential aspect of this empirical task is model selection, which is driven by the bias-variance tradeoff. This tradeoff captures the balance between a model's complexity and its generalizability to unseen data. High bias error reflects an overly simplistic model that fails to capture the underlying relationship between publication characteristics and citation outcomes, resulting in poor predictions. Both errors are also referred to as over and under fitting in statistical and machine learning literature \citep{hastie2009}.   

To balance the consequences from both errors, regularization techniques have been intensively investigated. In general, regularization refers to any set of techniques addressing over and under fitting. Particularly, reducing the effects of over fitting is achieved by making the model simpler. Conversely, letting the regularization techniques converge addresses the problems from under fitting \citep{james2013}. It has been noted, that regularization techniques are additionally accompanied by favorable properties. For instance, in high-dimensional data settings, the number of variables exceeds the number of observations making the estimation with the maximum likelihood principle infeasible. By imposing restrictions on the possible solutions, regularization can achieve viable estimation results. The problems of collinearity of variables in a design matrix can be similarly mitigated. Furthermore, choosing too many or too few variables can induce suboptimal predictions such that regularization provides an alternative estimation with build-in variable selection \citep{fahrmeier2013}.

Instead of depending exclusively on conventional metrics or scientometrician's judgment when choosing the best suitable prediction model, we aim to employ regularization techniques for the identification of important variables as well as additional robust checks. To this end, we propose to utilize variable selection in the framework of model-based gradient boosting (see, for example, \cite{mayr2014}) which, to the best of our knowledge, has not been applied in the context of citation prediction yet. Although the general performance is similar to classical regularization techniques, model-based gradient boosting has the advantage of a modular nature which allows for an alternative estimation of the regression function with a wide range of predictor effects in combination with inherent data-driven variable selection \citep{hepp2016}.

In the original domain of machine learning, boosting has been introduced in classification problems \citep{freund1995, freund1997}. Statistical boosting builds upon the machine learning algorithm by the additional consideration of the GLM framework. In this case, a loss function is expressed in terms of the negative log-likelihood (NLL), through which the outcome can be described in a similar fashion as in the GLM framework \citep{mccullagh1989, hastie2009, mayr2014}. In statistical boosting, the weak learners are referred to as base-learners which are combined to obtain a strong learner and there exists a wide range of possibilities to specify the form of these base-learners.
An additional extension is labeled model-based gradient boosting, which is based on the minimization of the empirical risk through steepest descent in function space \citep{friedman2001, hofner2014}. This allows for the estimation of the regression function via the specified loss function. In principle, the loss function can be again the NLL as in statistical boosting. Other options include the empirical risk or the residuals. The estimation is then based on evaluating the negative gradient of a loss function by fitting base-learners to the negative gradient \citep{buhlmann2003, mayr2014, hepp2016}.

The following algorithmic outline adapts the general model-based gradient boosting algorithm (see, for example, \cite{friedman2001,buhlman2007,mayr2014}) for the use-case in this work:
\begin{enumerate}
    \item Set $m = 0$. Choose offset values for the linear predictor $\hat{\eta}^{[0]} = (\mathbf{0})_{\{i = 1, \dots, n\}}$ and LM as base-learners, that is, $\beta_1x_1, \dots, \beta_px_p$.
    \item Update $m = m + 1$.
    \item Calculate the negative gradient vector $\mathbf{u}$ of the loss function $\rho(y_i, \eta)$ with respect to the linear predictor and plug in the current estimates $\hat{\eta}^{[m - 1]}$:
    \begin{equation*}
        \mathbf{u}^{[m]} = \left(u_{i}^{[m]}\right)_{i = 1, \dots, n} = \left(\frac{\partial}{\partial \eta} \rho(y_i, \eta) \Bigg|_{\eta = \hat{\eta}^{[m - 1]}} \right)
    \end{equation*}
    \item Fit the negative gradient vector $\mathbf{u}$ to each specified LM base-learner.
    \item Choose the base-learner $\hat{\beta}_{j^{*}}$ and component $j^{*}$ that fits the negative gradient vector based on the residual sum of squares best
    \begin{equation*}
        j^{*} = \operatorname*{argmin}_{1 \leq j \leq p} \sum_{i=1}^{n} \left(u^{[m]}_i - \hat{\beta}^{m}_{j}x_j\right)^2.
    \end{equation*}
    \item Utilize the component $j^{*}$ to update the linear predictor $\hat{\eta}$, such that
    \begin{equation*}
        \hat{\eta}^{[m]} = \hat{\eta}^{[m - 1]} + sl \cdot \hat{\beta}^{m}_{j^{*}} x_{j^{*}}
    \end{equation*}
    where $sl$ is a small learning rate or step-length.
    \item Continue steps (2) to (6) until specified stopping criterion $m = m_{\text{stop}}$ is reached.
\end{enumerate}

Model-based gradient boosting is not only a tool for an alternative estimation procedure to the established maximum likelihood principle, but also has an inherent data-driven variable selection mechanism. Recall that early stopping is the central approach to regularization in iterative machine learning algorithms. Since model-based gradient boosting is an iterative estimation algorithm, early stopping can be utilized to obtain a set of statistical models. Particularly, the scientometrician can immediately influence the performance by specifying an individual stopping criterion $m_{\text{stop}}$. As the main tuning parameter of the estimation algorithm, it can prevent over fitting and improve prediction accuracy. Convergence can be achieved by specifying a sufficiently large $m_{\text{stop}}$ such that the results are identical to the maximum likelihood estimation. Thus, it is the parameter controlling the bias-variance tradeoff for which variable selection and effects estimation shrinkage is additionally accounted \citep{mayr2012}.

Even though the performance of classical regularization techniques and model-based gradient boosting is comparable, the advantage of the latter is the modular nature which allows for the specification of a wide range of statistical models. Furthermore, there is a variety of loss functions and base-learners that can be utilized. It is also possible to do model choice by the quantification of the best base-learner for the variables. Consequently, model-based gradient boosting is a flexible and unifying framework for the alternative estimation of statistical models as well as regularization \citep{hepp2016}.

\subsection{Prediction and Data-driven Variable Selection via Model-based Gradient Boosting} \label{subsec:gbm}
In this section, the results for the alternative estimation procedure based on model-based gradient boosting with data-driven variable selection are presented. To this end, in the practical application the scientometrician has to decide the learning rate and the main tuning parameter $m_{\text{stop}}$. We set the learning rate for our data sets to $sl = 0.1$ since that is the usual practice \citep{hepp2016}. As stated before, the main tuning parameter $m_{\text{stop}}$ can in principle be chosen freely. In many real-world application settings, the number of parameters exceeds the number of observation such that it is often desirable to perform hyperparameter tuning.
Thus, cross-validation techniques including bootstrap, k-fold cross-validation or subsampling are usually utilized. The general idea is to iteratively apply performance evaluation to different subsamples of data such that over fitting is prevented by early stopping \citep{mayr2012}. We utilize 10-fold subsampling based on the train data sets. Afterward, the performance is evaluated based on the out-of-bag error, thereby performing hyperparameter tuning on an validation set. The optimal stopping criterion is then calculated as the average stopping criterion over all subsampling folds.Regarding the evaluation of the results, the performance criteria of the maximum likelihood models are retained. Additionally, the selection probabilities of the most important variable are reported. By doing so, we aim to capture the most important variables and their effect on the prediction of citations and mitigation of the Matthew effect. Results are reported for the boosted LM (GB-LM) and boosted negative binomial regression model (GB-NB).

\begin{table}[!htbp]
\centering
\caption{Ten most important variables in boosted LM and GLM with selection probabilities in brackets.} 
\label{tab:selopt} 
\begin{tabular}{cc} 
\\[-1.8ex] \hline
\hline \\[-1.8ex]
\textbf{GB-LM} & \textbf{GB-NB}  \\ 
\hline \\[-1.8ex]
H Score (0.073)    & Language 3 (0.059) \\ 
A Score (0.059)      & Age (0.056)       \\ 
References (0.051)      & References (0.053) \\ 
Age (0.046)      & Intercept (0.045)        \\ 
C Score (0.034)       & Research (0.034)     \\ 
Year 2017 (0.033)       & Year 2018 (0.033)     \\ 
Year 2016 (0.032)       & Year 2017 (0.031)     \\ 
Year 2015 (0.030)       & Year 2016 (0.030)     \\ 
Year 2014 (0.028)       & Year 2015 (0.029)     \\ 
Triangle C (0.027)       & Review (0.028)     \\ 
\hline \\[-1.8ex]
\end{tabular}
\end{table}

The results for the variable selection procedure with selection probabilities can be seen in Table \ref{tab:selopt}. Overall 89 out of 108 variable were chosen in GB-LM, out of which the animal and human proportions of the MeSH terms variables have the highest selection frequency. The MeSH terms are followed by the number of references and the average age of the referenced papers in terms of the selection probabilities. The weighted citations are further influenced by the year of the publication, particularly 2014 to 2017, although the selection frequencies are lower compared to the \textit{Score} variables. The selection probabilities can be best understood as the number of times the particular base-learner led to a reduction in the loss function across all iterations. For instance, the results show that the \textit{References} variable is chosen often across all boosting iterations, thereby highlight that the reduction in the loss function is in 5.1\% of iterations for this base-learner the best. The results in terms of frequently selected variable in GB-NB is comparable to the GB-LM counterpart. Generally, the \textit{References} variable is chosen similarly often, preceded by the average age of the references and the language of the paper. Furthermore, the years 2015 to 2018 are additionally important. The major difference is the non-existence of the proportion of the MeSH terms in the first ten most important variables and the high selection frequency of the language variable. Generally, the difference can partly be traced back to the different optimal stopping criteria for both models. The interpretation of the selection frequencies remains almost identical, although the reduction in the loss function is in 5.3\% of iterations for the \textit{References} base-learner the best.

\begin{table}[!htbp]
\centering
\caption{Estimated coefficients of boosted linear models (GB-LM) and boosted negative binomial regression (GB-NB) on respective train data sets. } 
\label{tab:opt} 
\begin{tabular}{lll} 
\\[-1.8ex] \hline
\hline \\[-1.8ex]
\textbf{Variable} & \textbf{GB-LM} & \textbf{GB-NB} \\ 
\hline \\[-1.8ex]
Clinical &  0.128 & 0.154   \\ 
Research &  0.113 &  0.180   \\ 
H Score & -0.303 &  \\ 
A Score &  -0.230 & -0.078  \\ 
C Score &  0.321 &  0.128  \\ 
Title & -0.036 &  -0.057 \\ 
Triangle ACH  &  0.192 & 0.103  \\ 
Triangle C  & -0.056 & -0.039   \\ 
Triangle H &  0.039 & -0.032   \\ 
Triangle Outside  &  0.159 &  0.066   \\ 
References &  0.707 &  0.566\\ 
Age & -0.504 & -0.243 \\ 
MeSH &  0.128 &  0.112 \\ 
Access &  0.065 &  0.256  \\ 
Language 2 & -0.807 & -0.568   \\ 
Language 3 &  -1.369 & -1.885 \\ 
Length &  0.063 & 0.143 \\ 
\hline
\hline
\end{tabular} 
\end{table}

In Table \ref{tab:opt} the results of the estimation coefficient for both models based on gradient boosting can be seen. Generally, the coefficients are in line with the results from the models estimated with maximum likelihood. For instance, the difference in absolute values between the coefficients for the \textit{References} variable between the full maximum likelihood model LMC and the GB-LM is $\Delta References = 0.707 - 0.713 = - 0.006$. Due to the modular nature of model-based gradient boosting, the interpretation of the effects of the \textit{References} variable remains identical as in the maximum likelihood approach. Thus, the increase in high number of references by 1\% increases the number of weighted citations only by 0.006\% less on average while holding the other variables constant. Likewise, an increase in low number of references by one unit increases the number of weighted citation only by 0.006 units less on average ceteris paribus compared to the full LMC. Similar coefficient differences are obtained in the negative binomial regression models, particularly when GB-NB is compared to the full maximum likelihood model GLMC. For instance, the difference in the \textit{References} variable is $\Delta References = 0.566 - 0.576 = - 0.010$. The interpretation remains due to the modular nature of model-based gradient boosting also identical for the negative binomial regression. In general, the results for the coefficient values are not surprising since the optimal stopping criterion for low-dimensional data settings is often large, such that the results are converging to the maximum likelihood estimator. On the same note, \cite{hepp2016} note that the exact maximum likelihood coefficients can be effectively recovered by imposing a sufficiently high stopping criterion. Furthermore, observe that the coefficients are not accompanied by standard errors or the corresponding p-values. Unfortunately, the nature of gradient boosting does not allow for classical statistical testing for significance of coefficient \citep{mayr2014}.

\begin{table}[!htbp]
\centering
\caption{Performance evaluation for the train and test data sets of boosted linear models (GB-LM) and boosted negative binomial regression (GB-NB).} 
\label{tab:perfopt} 
\begin{tabular}{lcccccc} 
\\[-1.8ex] \hline
\hline \\[-1.8ex]
& \multicolumn{2}{c}{\textbf{Train}} & \multicolumn{2}{c}{\textbf{Test}} \\ 
\cmidrule(r){2-3} \cmidrule(r){4-5} 
\textbf{Metric} & \textbf{GB-LM}  & \textbf{GB-NB} & \textbf{GB-LM} & \textbf{GB-NB}  \\ 
\hline \\[-1.8ex]
NLL & 16649688 & 45503820  & 4163489 & 11374516  \\ 
R$^{2}$ &  0.427 & -  & - & - \\
MSEP & 1.632 & 10575.638  & 1.632 & 15577.874  \\ 
MAE & 1.006 & 25.534  & 1.006 & 25.514  \\ 
\hline \\[-1.8ex]
\end{tabular}
\end{table}

In Table \ref{tab:perfopt} the results for the performance evaluation on train and test data can be seen. First, the GB-LM achieves a similar performance compared to the LMC model. The key differences is the parsimonious nature of the GB-LM, thereby achieving the results with 19 variables less than the LMC. In comparison to the LMI and LMR, the results indicate a much better performance for the GB-LM across all criteria. Generally, the largest improvement can be seen in the NLL which shows the lowest over all considered models for both test and train data sets. The only performance criterion where the LMC is better is the MSEP, although the differences is marginal. Severe differences are visible when the performance of the GLMC and the GB-NB are compared. Over all models considered, the GB-NB shows the best performance across all criteria for train and test data. Particularly, NLL, MSEP and MAE are all lower in GB-NB compared to the maximum likelihood counterpart models. This outcome highlights the general regularization behaviour of the model-based gradient boosting where consequences from the bias-variance tradeoff are mitigated.

The results indicate that model-based gradient boosting can effectively be utilized as an alternative estimation procedure with additional data-driven variable selection. Although the models were generally quite rich in terms of chosen variables, the general choice does not depend on preliminary and potential uncertain bias of the scientometrician, thereby providing a more neutral position of model selection. To show that more parsimonious models can be achieved by specifying a sufficiently low stopping criterion instead of relying on cross-validation techniques, while maintaining performance quality, additional results for these scenarios are presented in Appendix \ref{sec:low}.The results show again that a similar fit in terms of performance can be achieved with far less variables than considered in the large, non-parsimonious models. The key difference between classical regularization techniques and model-based gradient boosting lies in the flexibility of model selection. Unlike traditional methods that rely heavily on expertise or conservative criteria such as AIC and BIC in model choice, model-based gradient boosting offers a more adaptable approach. It allows for simultaneous estimation of coefficients and inherent data-driven variable selection, providing a more flexible alternative for model refinement.

\subsection{Results for Low Stopping Criteria in Model-based Gradient Boosting} \label{sec:low}
In section \ref{subsec:gbm}, cross-validation on subsamples is utilized to obtain the optimal stopping criterion. It has been noted, that in low-dimensional data settings, as it is the case for the citations data sets in this work, such a procedure is impractical since over fitting behaviour is generally slow \citep{buhlman2007}. Indeed, cross-validation shows that the optimal stopping criterion $m_{\text{stop}}$ is generally large, reaching values of around 3000 on average. Additionally, the number of coefficients at the optimal stopping criterion is much larger compared to the LMI and LMR. Since we are much more interested in the variable selection and the performance with as less direct involvement from the scientometrician as possible, we will manually set specific $m_{\text{stop}}$ criteria of 50 (50LM/50NB/), 100 (100LM/100NB) and 250 (250LM/250NB). Thus, we focus much more on variable selection and the performance on test data in comparison to the LM and GLM.

\begin{table}[!htbp]
\centering
\caption{Five most important variables in boosting of linear models with selection probabilities in brackets.} 
\label{tab:selprob} 
\begin{tabular}{lll} 
\\[-1.8ex] \hline
\hline \\[-1.8ex]
\textbf{50LM} & \textbf{100LM} & \textbf{250LM} \\ 
\hline \\[-1.8ex]
References (0.26)    & References (0.21)    & References (0.164)    \\ 
Year2018 (0.14)      & Year2018 (0.12)      & Age (0.136)    \\ 
Language3 (0.10)      & Age (0.12)           & Year2018 (0.088)   \\ 
Year2017 (0.08)      & Language3 (0.10)     & Year2017 (0.076)       \\ 
C.Score (0.08)       & Year2017 (0.09)      & Language3 (0.072) \\ 
\hline \\[-1.8ex]
\end{tabular}
\end{table}

In Table \ref{tab:selprob} the selection probabilities of the five most important variables after different stopping criteria $m_{\text{stop}}$ for the LM can be seen. In general, the absolute number of variables at different $m_{\text{stop}}$ is 12, 14 and 28, respectively. The results suggest that the most important variables is always the number of references. Afterward, usually the categorical variable for the year 2018 or the mean number of age of the cited paper. Furthermore, the language is also important indicating that the papers not in English carry important information on the weighted citations. Regarding the selection probabilities, the general tendency of decreasing proportion is not surprising since in each additional iteration a novel variable might be added to the model. Nevertheless, keeping in mind that the number of variables in the model after 250 iterations has increased almost three-fold in comparison to the model after 50, then it can be concluded that the proportion for the number of references has only decreased slowly.

\begin{table}[!htbp]
\centering
\caption{Performance evaluation for the train and test data sets of weighted citation with boosting of linear models.} 
\label{tab:perf3} 
\begin{tabular}{lcccccc} 
\\[-1.8ex] \hline
\hline \\[-1.8ex]
& \multicolumn{3}{c}{\textbf{Train}} & \multicolumn{3}{c}{\textbf{Test}} \\ 
\cmidrule(r){2-4} \cmidrule(r){5-7} 
\textbf{Metric} & \textbf{50LM} & \textbf{100LM} & \textbf{250LM} & \textbf{50LM} & \textbf{100LM} & \textbf{250LM} \\ 
\hline \\[-1.8ex]
NLL & 20307376 & 18810650 & 17384357 & 5080515 & 4706116 & 4348582 \\ 
R$^{2}$ & 0.301 & 0.353 & 0.402 & - & - & - \\
MSEP & 1.990 & 1.843 & 1.704 & 1.991 & 1.845 & 1.705 \\ 
MAE & 1.127 & 1.081 & 1.033 & 1.127 & 1.081 & 1.033 \\ 
\hline \\[-1.8ex]
\end{tabular}
\end{table}

Obviously, it is not only important to interpret the results for the variable selection but also to include information about the performance on the train and test data set. The results for the boosted LM can be seen in Table \ref{tab:perf3}. The results suggest that the NLL is slowly decrease for both train and test data with an increasing number of iterations. Thus, the model performs better with an increasing number of iterations which is also evident in the R$^{2}$ that is also increasing with the number of iterations. Consequently, the MSEP and MAE are thus decreasing with an increasing number of iterations. Comparing the absolute values of the performance criteria with the results obtained from the maximum likelihood LM, we conclude that the performance is quite similar. Actually, the final model with 32 variables comes very close the large, non-parsimonious model indicating that a similar fit can be obtained with a much simpler model.

\begin{table}[!htbp]
\centering
\caption{Five most important variables in boosting of binomial regression models with selection probabilities in brackets.} 
\label{tab:selprob2} 
\begin{tabular}{lll} 
\\[-1.8ex] \hline
\hline \\[-1.8ex]
\textbf{50NB} & \textbf{100NB} & \textbf{250NB} \\ 
\hline \\[-1.8ex]
References (0.70)      & References (0.38)    & Intercept (0.184)    \\ 
Intercept (0.18)      & Intercept (0.22)     & Language3 (0.172)    \\ 
Length (0.10)          & Language3 (0.22)     & References (0.156)   \\ 
Language3 (0.02)      & Length (0.11)        & Length (0.056)       \\ 
                     & MeSH (0.04)         & Journal Article (0.056) \\ 
\hline \\[-1.8ex]
\end{tabular}
\end{table}

Additionally, in Table \ref{tab:selprob2} the selection probabilities of the five most important variables after different stopping criteria $m_{\text{stop}}$ for the boosted GLM can be seen. Overall, the outcome is similar to the results from the boosted LM. Particularly, the results suggest that the most important variable is the number of references. Another common variable in the GLM and LM is the language which has a high selection probability in all models. The distinct differences between the outcomes lie in the variable with smaller selection probabilities. In the GLM, the length of the paper, number of MeSH terms as well as the dummy variable for a journal article play an important role whereas in the LM it is the year of the publication, proportion of cell MeSH terms and mean age of the references. Additionally, the intercept is chosen often in the GLM. In general, the absolute number of coefficients at different $m_{\text{stop}}$ is four, six and 20, respectively, which is also far less than in the LM. The magnitude of the selection probabilities is obviously driven again by the number of boosting iterations. The larger the iterations, the more novel variables are added to the final model, thereby reducing the proportion of the variables already added in earlier iterations.

\begin{table}[!htbp]
\centering
\caption{Performance evaluation for the train and test data sets of citation counts with boosting of negative binomial regression models.} 
\label{tab:perf4} 
\begin{tabular}{lcccccc} 
\\[-1.8ex] \hline
\hline \\[-1.8ex]
& \multicolumn{3}{c}{\textbf{Train}} & \multicolumn{3}{c}{\textbf{Test}} \\ 
\cmidrule(r){2-4} \cmidrule(r){5-7} 
\textbf{Metric} & \textbf{50NB} & \textbf{100NB} & \textbf{250NB} & \textbf{50NB} & \textbf{100NB} & \textbf{250NB} \\ 
\hline \\[-1.8ex]
NLL & 47253344 & 46736382 & 46149896 & 11810862 & 11686199 & 11555123 \\ 
MSEP &  10885.980 & 10839.241 & 10743.703 & 15887.852 & 15841.734 & 15749.540 \\ 
MAE & 29.256 & 28.092 & 26.514 & 29.230 & 28.061 & 26.488 \\ 
\hline \\[-1.8ex]
\end{tabular}
\end{table}

The performance evaluation for the boosting of GLM can be seen in Table \ref{tab:perf4}. In general, we conclude that all performance criteria are comparable to the result from the GLM. The NLL is again slowly decreasing with the number of iterations indicating an improvement in the fit. Similarly, MSEP and MAE are also decreasing. Comparing the performance from the maximum likelihood GLM and the boosted GLM, the results show a comparable performance on train and test data. Particularly, the MSEP and MAE of the boosting models come very close the large, non-parsimonious GLM. 
\end{appendices}

\bibliography{sn-bibliography}

\end{document}